\documentclass[letterpaper, 10 pt, conference]{ieeeconf}  

\IEEEoverridecommandlockouts                              

\usepackage{cite}
\usepackage{amsmath,amssymb,amsfonts}
\usepackage{algorithmic}
\usepackage{graphicx}
\usepackage{textcomp}
\usepackage{xcolor}
\usepackage{booktabs}
\usepackage{pifont}
\usepackage{hyperref}
\usepackage{colortbl}
\definecolor{darkgreen}{rgb}{0.0, 0.7, 0.0} 
\hypersetup{
    colorlinks=true,        
    linkcolor=black,        
    citecolor=black,        
    filecolor=black,        
    urlcolor=blue           
}

\title{\LARGE \bf
Lightweight ResNet-Based Deep Learning for Photoplethysmography Signal Quality Assessment
}

\author{Yangyang Zhao$^{1*}$, Matti Kaisti$^{1}$, Olli Lahdenoja$^{1}$, Jonas Sandelin$^{1}$, Arman Anzanpour$^{1}$,\\
        Joonas Lehto$^{2}$, Joel Nuotio$^{2}$, Jussi Jaakkola$^{2}$,
        Arto Relander$^{2}$, Tuija Vasankari$^{2}$, \\
        Juhani Airaksinen$^{2}$, Tuomas Kiviniemi$^{2}$, Tero Koivisto$^{1}$%
\thanks{This study was funded by Moore4Medical project which received funding from the ECSEL JU and Business Finland, 
        under grant agreement H2020-ECSEL-2019-IA-876190 and 7215/31/2019.}%
\thanks{$^{1}$Y. Zhao, M. Kaisti, O. Lahdenoja, J. Sandelin, A. Anzanpour, and T. Koivisto are with the Department of Computing, 
        Faculty of Technology, University of Turku, Turku 20520, Finland 
        (e-mail: yazhao@utu.fi; mkaist@utu.fi; olanla@utu.fi; jojusan@utu.fi; armanz@utu.fi; tejuko@utu.fi).}%
\thanks{$^{2}$J. Lehto, J. Nuotio, J. Jaakkola, A. Relander, T. Vasankari, J. Airaksinen, and T. Kiviniemi 
        are with the Heart Center, Turku University Hospital, Turku 20521, Finland 
        (e-mail: jojuleh@utu.fi; joel.nuotio@utu.fi; jussi.jaakkola@utu.fi; arnire@utu.fi; 
        tuija.vasankari@tyks.fi; juhani.airaksinen@tyks.fi; tuomas.kiviniemi@tyks.fi).}%
\thanks{*Corresponding author: Yangyang Zhao (e-mail: yazhao@utu.fi)}%
}

\begin{document}

\maketitle
\begin{abstract}

With the growing application of deep learning in wearable devices, lightweight and efficient models are critical to address the computational constraints in resource-limited platforms. The performance of these approaches can be potentially improved by using various preprocessing methods.
This study proposes a lightweight ResNet-based deep learning framework with Squeeze-and-Excitation (SE) modules for photoplethysmography (PPG) signal quality assessment (SQA) and compares different input configurations, including the PPG signal alone, its first derivative (FDP), its second derivative (SDP), the autocorrelation of PPG (ATC), and various combinations of these channels.
Experimental evaluations on the Moore4Medical (M4M) and MIMIC-IV datasets demonstrate the model's performance, achieving up to 96.52\% AUC on the M4M test dataset and up to 84.43\% AUC on the MIMIC-IV dataset. The novel M4M dataset was collected to explore PPG-based monitoring for detecting atrial fibrillation (AF) and AF burden in high-risk patients. 
Compared to the five reproduced existing studies, our models achieves over 99\% reduction in parameters and more than 60\% reduction in floating-point operations (FLOPs). 

{\textbf{\textit{Clinical Relevance}}}\textemdash Accurate PPG signal quality assessment is crucial for continuous cardiovascular monitoring. By reducing false alarms and enhancing detection reliability, the proposed lightweight framework supports clinical decisions and practical deployment in resource-limited wearable devices, aiding broader adoption in telemedicine and remote care.
\end{abstract}

\section{INTRODUCTION}

Photoplethysmography (PPG) signals are widely used in wearable devices to measure physiological parameters such as heart rate \cite{motin2017ensemble} and blood pressure \cite{pereira2020photoplethysmography}, or to detect cardiac conditions \cite{stehlik2020continuous} \cite{zhang2022refined}.
However, the quality of PPG signals can be degraded by motion artifacts, noise, and poor sensor-skin contact, resulting to inaccuracies in downstream analyses \cite{orphanidou2018quality}. Signal quality assessment (SQA) is essential for identifying low-quality signals and ensuring the reliability of wearable health monitoring.

Deep learning-based approaches have emerged as a solution to the limitations of traditional methods, offering automatic feature extraction and higher adaptability. For instance, one-dimensional (1D) models PPG signals directly, utilizing convolutional neural networks (CNNs) to classify signal quality \cite{naeini2019real}\cite{shin2022deep}. On the other hand, two-dimensional (2D) models transform PPG signals into image representations, such as recurrence plots or time-frequency images, and use architectures like ResNet and VGG for classification \cite{roh2021recurrence}\cite{chen2021signal}. However, these deep learning-based methods, particularly those involving image transformations, often require substantial computational resources and large numbers of parameters, making them less practical for resource-constrained devices like wearables.

To mitigate this limitation, this study proposes a lightweight ResNet-based framework enhanced with channel attention (Squeeze-and-Excitation, SE) modules for efficient SQA. 
We compare different inputs, including the raw PPG signal, its first derivative (FDP), second derivative (SDP), autocorrelation of PPG signal, and various combinations of these signals. This lightweight design eliminates signal-to-image conversion, reducing computational complexity and parameter count while efficiently capturing key temporal and dynamic characteristics. We also conduct a comparison with existing methods to evaluate the effectiveness of the proposed approach.

This study is organized as follows:
Section II (Dataset) outlines the datasets and quality labeling rules.
Section III (Methods) describes the processing, segmentation, and ResNet-based model architecture, as well as the training settings, experimental setup, and evaluation metrics.
Section IV (Experimental Results) presents the ablation study, comparisons with baseline models and related studies.
The discussion and conclusion are presented in Sections V and VI, respectively.

\section{Dataset}

This study used two datasets for model development and evaluation: the Moore4Medical (M4M) dataset and the MIMIC-IV Waveform dataset.

The M4M dataset originally included PPG measurements from 49 participants collected in a hospital setting between September 2022 and August 2023. However, data from one participant were not included in the analysis due to unavailability during dataset preparation, resulting in a final dataset of 48 participants. Among these, 34 participants had sinus rhythm (SR), and 14 had atrial fibrillation (AF). PPG signals were recorded during long-term monitoring using a wrist-worn Philips Datalogger with reflective green LEDs using sampling frequency of 32 Hz, as illustrated in \textbf{Fig. 1}.

The recordings were divided into two subsets: the M4M train dataset and the M4M test dataset. The training set included 41 participants (23 females and 18 males) with a mean age of 69.7 years, height of 169.3 cm, and weight of 83.9 kg, while the test set comprised 7 participants (all males) with a mean age of 66.1 years, height of 178.6 cm, and weight of 98.5 kg.

The M4M train dataset consists of 774.1 hours of recordings, with 58.5\% labeled as good quality, while the test dataset includes 159.1 hours of recordings, with 52.1\% labeled as good quality. The test dataset contains the last seven individuals recruited during data collection, who were excluded from the training set.

The MIMIC-IV Waveform dataset was used exclusively as an additional independent test set and includes 14 randomly selected PPG records, representing 83.2 hours of data with 72.1\% labeled as good quality. These records were sampled from the publicly available MIMIC-IV database (version 0.1.0), which contains de-identified physiological signals collected from critically ill patients at the Beth Israel Deaconess Medical Center \cite{moody2022mimiciv}\cite{goldberger2000physiobank}.

The PPG signals were annotated manually by clinical experts with cardiology training, following a standardized protocol based on signal morphology, noise level, and beat visibility. Visual inspection was performed independently by two experts, and discrepancies were resolved through consensus to ensure consistency and accuracy. Due to the time-consuming nature of manual annotation, only 14 high-quality PPG records from the MIMIC-IV database were selected for validation purposes. \textbf{Fig. 2} illustrates representative examples of good- and bad-quality segments from the M4M dataset.

\section{ Methods}

\subsection{ Pipeline}

\textbf{Fig. 1} illustrates the proposed platform for the PPG SQA workflow, including data collection and processing methods (bandpass filtering, normalization and segmentation), training and validation strategies, and the final testing stage. 

\begin{figure*}[htbp] 
       \centering 
       \includegraphics[width=\textwidth]{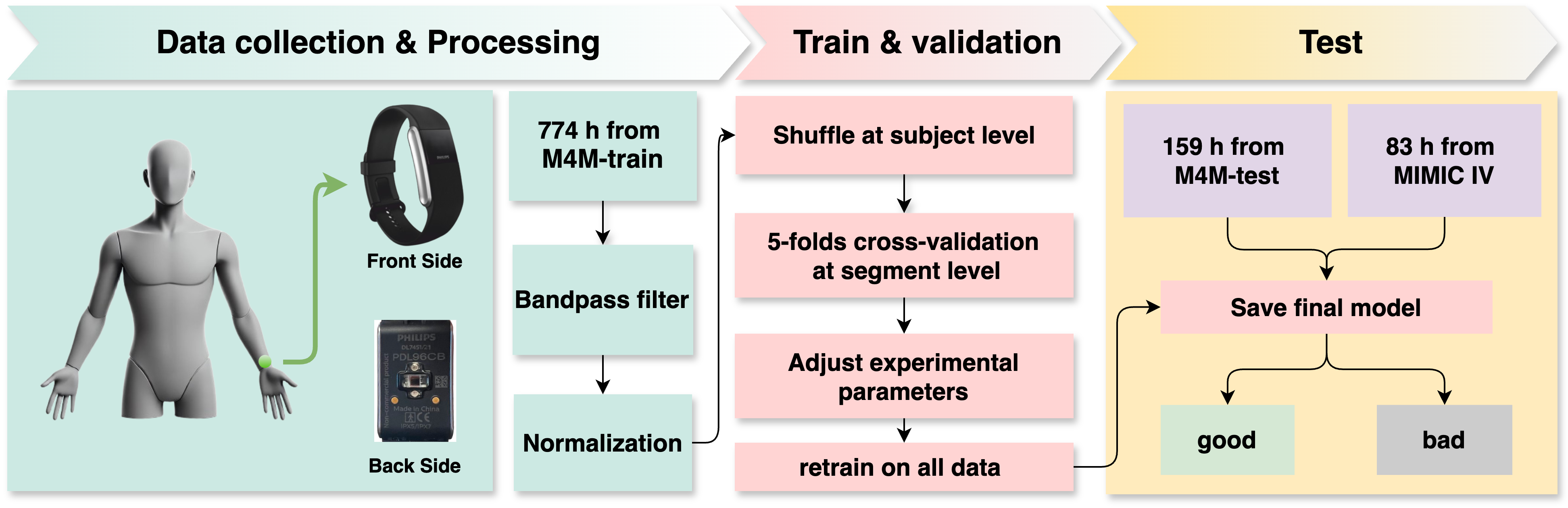} 
       \caption{Pipeline for PPG signal quality assessment. Data from the M4M train dataset undergoes preprocessing (bandpass filtering and normalization) and 5-fold cross-validation on train data to adjust parameters. The model is then retrained on the entire M4M train dataset and tested on the M4M test and MIMIC-IV datasets, classifying signals as good or bad.} 
       \label{fig1} 
\end{figure*}

\subsection{Pre-processing and Segmentation}

We applied a third-order Butterworth bandpass filter (0.5–8 Hz) to eliminate noise and applied normalization, setting the mean to zero and standard deviation to one, to prepare the signals for analysis. The filtered signals were divided into non-overlapping 30-second segments, with each segment labeled as good if over 80\% of the signal within it was of good quality. \textbf{Fig. 2} illustrates examples of raw signals labeled as good and bad quality, along with their processed signal (Clean), FDP, SDP, and autocorrelation of PPG signal (ATC), which is included for additional comparison.

\begin{figure}[htbp] 
       \centering 
       \includegraphics[width=\columnwidth]{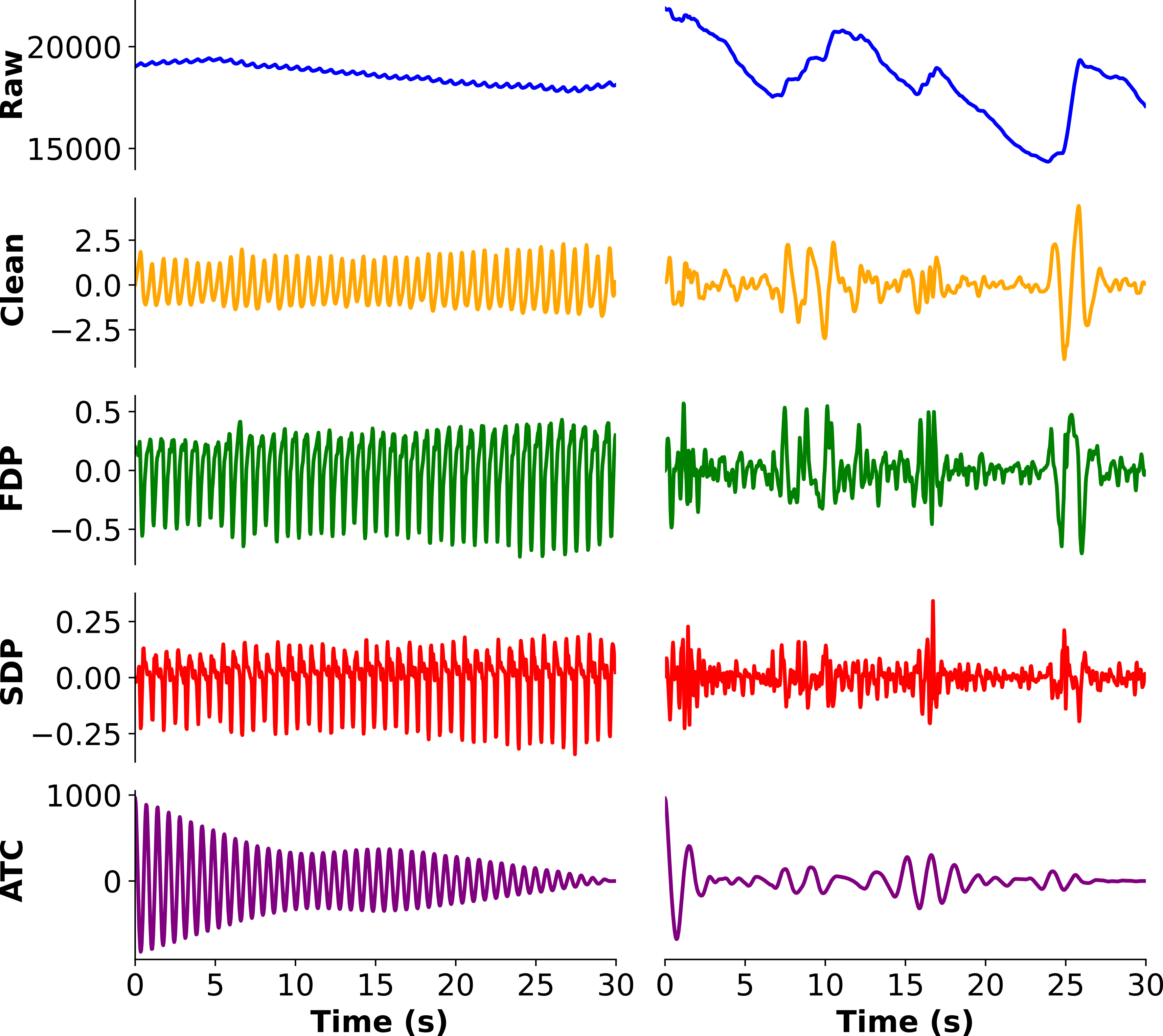} 
       \caption{Comparison of good quality (left) and bad quality (right) PPG signals over a 30-second segment. The rows represent the raw signal, preprocessed signal (Clean), first derivative of PPG (FDP), second derivative of PPG (SDP), and autocorrelation of PPG (ATC). The y-axis indicates the amplitude for each signal type.} 
       \label{fig2} 
    \end{figure}

\subsection{Model Architecture}

The architecture integrates residual learning, channel attention, and global feature aggregation. As illustrated in \textbf{Fig. 3}, the proposed framework incorporates detailed components, including Res blocks and Basic blocks with integrated SE modules for channel attention.
The code is available on \href{https://github.com/lady052888/A-lightweight-Resnet-with-SE-block}{GitHub}

\begin{figure*}[htbp] 
       \centering 
       \includegraphics[width=\textwidth]{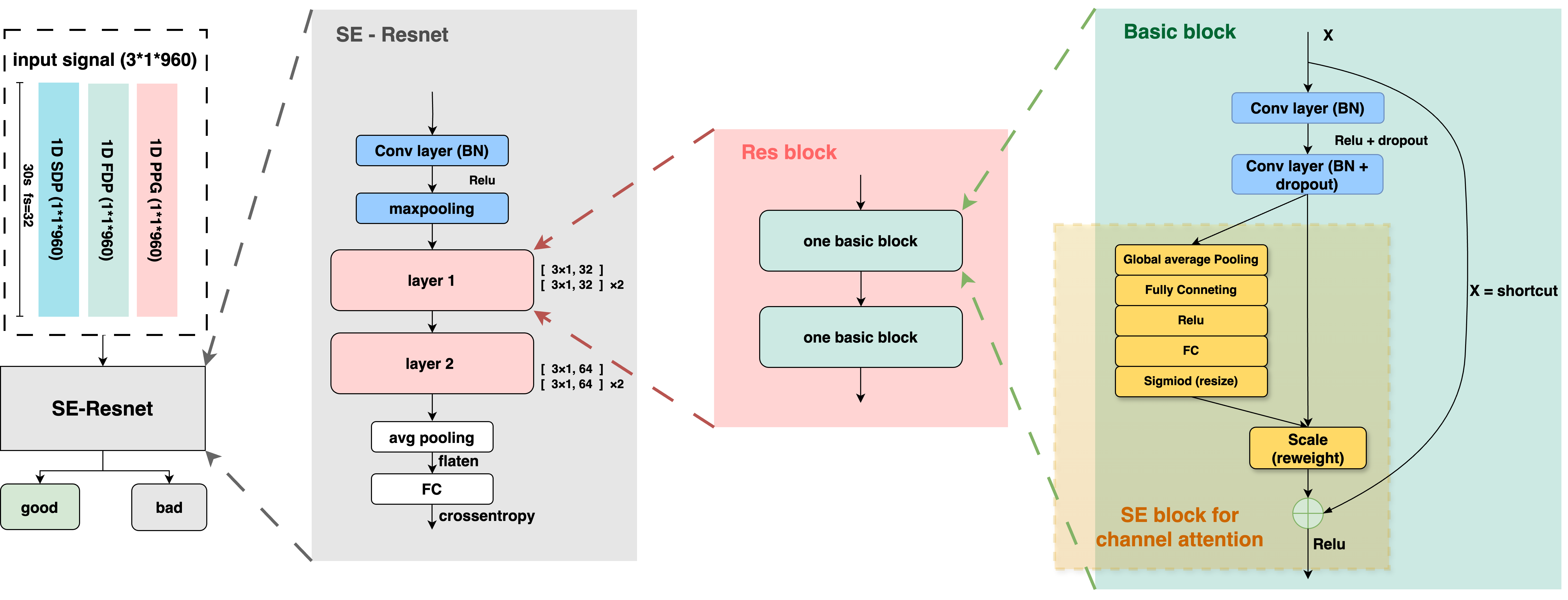} 
       \caption{Overview of the proposed LRS-SE framework. The first panel illustrates the input signal format, comprising three channels (the raw PPG signal, the first derivative of PPG (FDP) and the second derivative of PPG (SDP)) and the overall LRS-SE architecture for classifying signal quality as good or bad. The second panel shows the detailed structure of LRS-SE, which includes two Res blocks (Layer 1 and Layer 2). The third panel provides the composition of a single Res block, consisting of two Basic blocks. The fourth panel presents the structure of a Basic block, which integrates a Squeeze-and-Excitation (SE) block to implement channel attention.}
 
       \label{fig3} 
    \end{figure*}
    
\subsubsection{Residual Learning with Basic Blocks}

The network consists of two residual stages, each containing stacked BasicBlock1D modules with shortcut connections \cite{he2016deep}, as shown in \textbf{Fig. 3}. The first stage uses convolutional layers with 32 filters, and the second stage increases the number of filters to 64. Each BasicBlock1D module consists of two convolutional layers with 3×3 filters, stride 1, and padding 1, followed by Batch Normalization (BN) and ReLU activation. Dropout layers with a probability of $p=0.2$ are applied after each convolutional layer.

\subsubsection{Channel Attention Mechanism}

A Squeeze-and-Excitation (SE) module \cite{hu2018squeeze} is incorporated after each BasicBlock1D module, as highlighted in the yellow section of \textbf{Fig. 3}. The SE module recalibrates channel-wise features by leveraging global context, which is computed through adaptive average pooling, followed by two fully connected layers and a sigmoid activation. The intermediate dimension during the recalibration process is determined by a reduction ratio $r$, empirically set to 8 based on experimental results for optimal performance. This mechanism models interdependencies between channels by adaptively recalibrating their feature response strengths, guided by the global context of the input signal.

\subsubsection{Global Representation and Regularization}

The network begins with a convolutional layer (kernel size: 7, stride: 2, padding: 3) with 32 filters, followed by Batch Normalization, a ReLU activation, and a max-pooling layer (kernel size: 3, stride: 2). The input to the network is a three-channel signal with a shape of [64, 3, 960], where 960 corresponds to a 30-second segment sampled at 32 Hz, and the batch size is set to 64. After passing through the residual and attention blocks, an Adaptive Average Pooling (AAP) layer aggregates features across the temporal dimension. The pooled features are then fed into a fully connected (FC) layer to produce class probabilities.

\subsection{Experimental Settings}

\subsubsection{Training setting}

We employ a two-stage training strategy involving 5-fold cross-validation followed by full training on the entire training dataset. The M4M training data is first shuffled at the subject level and split into five subsets. In each fold, four subsets are used for training and the remaining one for validation. To enhance training diversity, signal segments within each training fold are reshuffled at the start of every epoch.

To ensure reproducibility, random seeds were fixed across Python’s random module, NumPy, and PyTorch. Fold-specific randomness was introduced by assigning a unique seed based on a combination of a global seed and the fold index.

After the optimal model configuration was identified through cross-validation, the model was retrained from scratch using the entire M4M training dataset, which combines all training and validation subjects. During this stage, signal segments were globally shuffled before each epoch. The resulting model was saved as the final model and evaluated on the independent test set to assess its generalization performance.

Model training employed the Adam optimizer with a learning rate of \(1 \times 10^{-4}\) and a weight decay of \(1 \times 10^{-5}\), facilitating efficient convergence while mitigating overfitting. A StepLR learning rate scheduler was used to reduce the learning rate by a factor of 0.1 every 20 epochs, over a total of 60 epochs. The binary classification task was optimized using the CrossEntropyLoss function.

\subsubsection{Experimental Environment}

Model training was carried out in a Linux-based environment using Python 3.11.9 and the PyTorch deep learning framework (Version 2.3.0). An NVIDIA TITAN RTX graphics card with 23.64 GB of VRAM was used for GPU acceleration, with CUDA 12.1 enabling parallel computation.

\subsubsection{Evaluation Metrics}

Model performance was evaluated using three metrics: the Area Under the Receiver Operating Characteristic Curve (AUC), the total number of model parameters, and the number of Floating-Point Operations (FLOPs). AUC quantifies the model's classification performance across all decision thresholds. The number of parameters reflects the model’s representational capacity, while FLOPs estimate the computational cost associated with a single forward pass during inference.

\section{EXPERIMENTAL RESULTS}

\subsection{Ablation Study of the Proposed Network Architecture}

The impact of different input configurations is evaluated in the first experiment. 
The inputs include single-channel data, such as the PPG signal, FDP, SDP, and ATC, as well as various combinations of these signals. 
The results are evaluated on the M4M test set and MIMIC datasets. The results are further categorized into two groups: models without the SE block and models with the SE block.

\textbf{Table I} presents the results of different input combinations, including comparisons of AUC, parameter count, and FLOPs, with green and blue highlighting the deviations (increases or decreases) relative to the mean for each column.
Without the SE block, the AUC on the M4M test set varied from -0.19\% to +0.12\%, indicating that increasing the number of inputs had no significant impact on performance improvement. 
However, on the MIMIC dataset, the variation ranged from -4.79\% to +2.61\%, reflecting a certain degree of fluctuation.
Among these, the use of ATC as an input showed the poorest performance, with combinations involving ATC demonstrating a mean decrease of -1.38\% relative to the average. In contrast, PPG+FDP, PPG+SDP, and PPG+FDP+SDP exhibited better performance, with an average improvement of 1.92\%.
With the SE block, the AUC on the M4M test set ranged from -0.29\% to +0.25\%, with ATC as the input achieving the best performance. Other combinations, including three- and four-signal inputs, showed consistent improvements, with an average AUC of 0.9641, representing an increase of approximately 0.14\%.
On the MIMIC dataset, the variation ranged from -14.26\% to +8.54\%. The optimal performance was achieved with PPG as a single input, showing an improvement of 8.54\%, followed by PPG+FDP+SDP (6.58\% improvement).
The use of the SE block increased the mean AUC on the M4M test set by 0.25\%, but decreased the mean AUC on the MIMIC dataset by 6.75\%. Moreover, the SE block only led to minor increases in parameter count (2.32k) and MMAC (0.05 MMAC), with negligible impact on computational complexity.

Overall, PPG+FDP+SDP with SE demonstrated the most stable and robust performance. Additionally, the four-signal combination with SE showed similar performance to the optimal combination but with limited improvement. PPG alone with SE also performed well.

\begin{table*}[htbp] 
\renewcommand{\arraystretch}{1.5}
\centering 
\caption{Performance Comparison Across Methods on the M4M test set, showing deviations from the mean for each column.} 
\resizebox{\textwidth}{!}{%
\begin{tabular}{lcccccccc} 
\hline 
\textbf{} & \multicolumn{4}{c}{\textbf{NO-SE}} & \multicolumn{4}{c}{\textbf{with-SE}} \\
\cline{2-5} \cline{6-9}
\textbf{Input} & \textbf{M4M (AUC)} & \textbf{MIMIC (AUC)} & \textbf{Params (k)} & \textbf{FLOPs (MMAC)} & \textbf{M4M (AUC)} & \textbf{MIMIC (AUC)} & \textbf{Params (k)} & \textbf{FLOPs (MMAC)} \\ 
\hline 
PPG            & 0.9604 \textcolor{darkgreen}{- \textbf{0.00\%}} & 0.8253 \textcolor{darkgreen}{$\uparrow \textbf{0.30\%}$} & 58.66 & 8.67 & 0.9617 \textcolor{blue}{$\downarrow \textbf{0.11\%}$} & 0.8328 \textcolor{darkgreen}{$\uparrow \textbf{8.54\%}$} & 61.22 & 8.7 \\
FDP            & 0.9611 \textcolor{darkgreen}{$\uparrow \textbf{0.07\%}$} & 0.8213 \textcolor{blue}{$\downarrow \textbf{0.18\%}$} & 58.66 & 8.67 & 0.9624 \textcolor{blue}{$\downarrow \textbf{0.04\%}$} & 0.8043 \textcolor{darkgreen}{$\uparrow \textbf{4.82\%}$} & 61.22 & 8.7 \\
SDP            & 0.9598 \textcolor{blue}{$\downarrow \textbf{0.06\%}$} & 0.8312 \textcolor{darkgreen}{$\uparrow \textbf{1.02\%}$} & 58.66 & 8.67 & 0.9620 \textcolor{blue}{$\downarrow \textbf{0.08\%}$} & 0.7922 \textcolor{darkgreen}{$\uparrow \textbf{3.25\%}$} & 61.22 & 8.7 \\
ATC            & 0.9615 \textcolor{darkgreen}{$\uparrow \textbf{0.11\%}$} & 0.7834 \textcolor{blue}{$\downarrow \textbf{4.79\%}$} & 58.66 & 8.67 & 0.9652 \textcolor{darkgreen}{$\uparrow \textbf{0.25\%}$} & 0.7139 \textcolor{blue}{$\downarrow \textbf{6.96\%}$} & 61.22 & 8.7 \\
PPG+FDP        & 0.9596 \textcolor{blue}{$\downarrow \textbf{0.08\%}$} & 0.8348 \textcolor{darkgreen}{$\uparrow \textbf{1.46\%}$} & 58.88 & 8.78 & 0.9600 \textcolor{blue}{$\downarrow \textbf{0.29\%}$} & 0.8096 \textcolor{darkgreen}{$\uparrow \textbf{5.51\%}$} & 61.44 & 8.81 \\
PPG+SDP        & 0.9585 \textcolor{blue}{$\downarrow \textbf{0.19\%}$} & 0.8443 \textcolor{darkgreen}{$\uparrow \textbf{2.61\%}$} & 58.88 & 8.78 & 0.9601 \textcolor{blue}{$\downarrow \textbf{0.28\%}$} & 0.7986 \textcolor{darkgreen}{$\uparrow \textbf{4.08\%}$} & 61.44 & 8.81 \\
PPG+ATC        & 0.9616 \textcolor{darkgreen}{$\uparrow \textbf{0.12\%}$} & 0.8167 \textcolor{blue}{$\downarrow \textbf{0.74\%}$} & 58.88 & 8.78 & 0.9631 \textcolor{darkgreen}{$\uparrow \textbf{0.03\%}$} & 0.7308 \textcolor{blue}{$\downarrow \textbf{4.76\%}$} & 61.44 & 8.81 \\
PPG+FDP+SDP    & 0.9603 \textcolor{blue}{$\downarrow \textbf{0.01\%}$} & 0.8368 \textcolor{darkgreen}{$\uparrow \textbf{1.70\%}$} & 59.11 & 8.89 & 0.9638 \textcolor{darkgreen}{$\uparrow \textbf{0.10\%}$} & 0.8178 \textcolor{darkgreen}{$\uparrow \textbf{6.58\%}$} & 61.67 & 8.92 \\
PPG+FDP+ATC    & 0.9612 \textcolor{darkgreen}{$\uparrow \textbf{0.08\%}$} & 0.8233 \textcolor{darkgreen}{$\uparrow \textbf{0.06\%}$} & 59.11 & 8.89 & 0.9643 \textcolor{darkgreen}{$\uparrow \textbf{0.16\%}$} & 0.6747 \textcolor{blue}{$\downarrow \textbf{12.07\%}$} & 61.67 & 8.92 \\
PPG+SDP+ATC    & 0.9611 \textcolor{darkgreen}{$\uparrow \textbf{0.07\%}$} & 0.8164 \textcolor{blue}{$\downarrow \textbf{0.78\%}$} & 59.11 & 8.89 & 0.9644 \textcolor{darkgreen}{$\uparrow \textbf{0.17\%}$} & 0.6579 \textcolor{blue}{$\downarrow \textbf{14.26\%}$} & 61.67 & 8.92 \\
PPG+FDP+SDP+ATC & 0.9593 \textcolor{blue}{$\downarrow \textbf{0.11\%}$} & 0.8174 \textcolor{blue}{$\downarrow \textbf{0.66\%}$} & 59.33 & 8.99 & 0.9640 \textcolor{darkgreen}{$\uparrow \textbf{0.12\%}$} & 0.8081 \textcolor{darkgreen}{$\uparrow \textbf{5.37\%}$} & 61.89 & 9.03 \\
\rowcolor[gray]{0.9} 
\midrule
\textbf{Mean}  & \textbf{0.9604} & \textbf{0.8228} & \textbf{58.90} & \textbf{8.79} & \textbf{0.9628} & \textbf{0.7673} & \textbf{61.22} & \textbf{8.84} \\
\hline 
\end{tabular}%
} 
\end{table*}

\subsection{Comparison with Baseline Models}

\textbf{Table II} compares the proposed method with existing approaches in key performance metrics and model complexity, including parameter count and FLOPs.

To ensure a fair comparison, the referenced methods were reproduced as part of this study. Some network parameters, such as strides, padding, and input dimensions, were not detailed in the original papers; therefore, standard practices and widely accepted conventions were applied to maintain reproducibility. All reproductions used consistent settings, including 30-second segments, a batch size of 64, and uniform preprocessing.

Specifically, Naeini \emph{et al.} \cite{naeini2019real} implemented a single convolutional layer with 196 filters (1×16), ReLU activation, max pooling (1×4), and a fully connected layer with 1024 neurons and Softmax, applying Dropout (0.05) and L2 regularization to mitigate overfitting.  
Goh \emph{et al.} \cite{goh2020robust} utilized a 2-layer convolutional network with 50 filters (1×50) per layer, ReLU activation, max pooling (1×5), and Dropout (0.5), followed by a fully connected layer with 500 neurons.  
Shin \emph{et al.} \cite{shin2022deep} employed a 6-layer convolutional network with multi-scale kernels (10×1 to 2×1), batch normalization, max pooling, and Dropout (0.2). A fully connected layer with 1546 neurons was used for classification.  
Only the 4-layer 1D-CNN from Sivanjaneyulu \emph{et al.} \cite{sivanjaneyulu2022cnn} was reproduced, utilizing filters (16, 32, 64) of size 3×1 with Batch Normalization, max pooling (3×3), and Dropout (0.2), followed by fully connected layers (128$\rightarrow$64$\rightarrow$32$\rightarrow$1).  
Finally, Naeini \emph{et al.} \cite{naeini2023deep} employed a 1D CNN with two convolutional layers (32 and 64 filters), batch normalization, max pooling, and a fully connected layer with 512 neurons and sigmoid activation.  

The comparison in \textbf{Table II}  shows that the proposed lightweight network achieves performance improvements regardless of whether PPG is used as a single input or in combination with other inputs.
On the M4M test set, the AUC improvements for PPG+FDP+SDP\_SE and PPG+FDP+SDP+ATC\_SE exceed 1.45\%, specifically 1.46\% and 1.48\%, respectively.
On the MIMIC dataset, PPG+FDP+SDP achieves the highest improvement (8.86\%), while the addition of the SE block with the same input yields a 6.39\% improvement.

\begin{table}[htbp]
\renewcommand{\arraystretch}{1.3}
\centering
\caption{Performance Comparison Across Methods, showing deviations from the mean for each column relative to the reproduced SOTA model.}
\begin{tabular}{p{1.5cm}p{1.3cm}p{1.3cm}p{1.3cm}p{1.3cm}}
\hline
\textbf{Methods} & \textbf{M4M (AUC)} & \textbf{Mimic (AUC)} & \textbf{Params (M)} & \textbf{FLOPs (MMAC)} \\
\hline
2019\cite{naeini2019real} & 0.9468 & 0.7409 & 43.37 & 50.7 \\
2020\cite{goh2020robust} & 0.9556 & 0.7959 & 1.42 & 15.93 \\
2022\cite{shin2022deep} & 0.9399 & 0.7819 & 6.71 & 23.58 \\
2022\cite{sivanjaneyulu2022cnn} & 0.9534 & 0.7575 & 1.33 & 10.78 \\
2023\cite{naeini2023deep} & 0.9536 & 0.7671 & 10.7 & 17.02 \\
\midrule
Mean & 0.9499 & 0.7687 & 12.306 & 23.602 \\
\midrule
PPG\_SE & 0.9617 & 0.8328 & 0.0612 & 8.70 \\
& \cellcolor[gray]{0.9}\textcolor{darkgreen}{$\uparrow \textbf{1.24\%}$} & \cellcolor[gray]{0.9}\textcolor{darkgreen}{$\uparrow \textbf{8.34\%}$} & \cellcolor[gray]{0.9}\textcolor{blue}{$\downarrow \textbf{99.50\%}$} & \cellcolor[gray]{0.9}\textcolor{blue}{$\downarrow \textbf{63.14\%}$} \\
\midrule
PPG+FDP & 0.9603 & 0.8368 & 0.0591 & 8.89 \\
+SDP & \cellcolor[gray]{0.9}\textcolor{darkgreen}{$\uparrow \textbf{1.09\%}$} & \cellcolor[gray]{0.9}\textcolor{darkgreen}{$\uparrow \textbf{8.86\%}$} & \cellcolor[gray]{0.9}\textcolor{blue}{$\downarrow \textbf{99.52\%}$} & \cellcolor[gray]{0.9}\textcolor{blue}{$\downarrow \textbf{62.33\%}$} \\
\midrule
PPG+FDP & 0.9638 & 0.8178 & 0.0618 & 8.92 \\
+SDP\_SE & \cellcolor[gray]{0.9}\textcolor{darkgreen}{$\uparrow \textbf{1.46\%}$} & \cellcolor[gray]{0.9}\textcolor{darkgreen}{$\uparrow \textbf{6.39\%}$} & \cellcolor[gray]{0.9}\textcolor{blue}{$\downarrow \textbf{99.50\%}$} & \cellcolor[gray]{0.9}\textcolor{blue}{$\downarrow \textbf{62.21\%}$} \\
\midrule
PPG+FDP+ & 0.9640 & 0.8081 & 0.0619 & 9.03 \\
SDP+ATC\_SE & \cellcolor[gray]{0.9}\textcolor{darkgreen}{$\uparrow \textbf{1.48\%}$} & \cellcolor[gray]{0.9}\textcolor{darkgreen}{$\uparrow \textbf{5.13\%}$} & \cellcolor[gray]{0.9}\textcolor{blue}{$\downarrow \textbf{99.50\%}$} & \cellcolor[gray]{0.9}\textcolor{blue}{$\downarrow \textbf{61.74\%}$} \\
\hline
\end{tabular}
\end{table}

\subsection{Comparison with Results from Related Studies}

The \textbf{Table III} summarizes the performance of the proposed method in comparison with several state-of-the-art (SOTA) approaches reported in the literature. These methods span a variety of datasets, input representations, and experimental setups, providing a comprehensive benchmark for evaluating the effectiveness of SQA techniques.

\begin{table*}[htbp]
\renewcommand{\arraystretch}{1.5}
\centering
\caption{Comparison With Other State-of-the-Art Works. (The bolded in grey background references correspond to works that were reproduced in Experiment II for comparison. Duration values are approximate results calculated based on descriptions in the original paper and are for reference only. "-" indicates that the original paper did not provide the relevant information.)}
\label{tab:comparison}
\resizebox{\textwidth}{!}{%
\begin{tabular}{p{1.3cm}p{2.8cm}p{1.3cm}p{1cm}p{4.5cm}p{3cm}p{1cm}p{1cm}}
\hline
\textbf{Year} & \textbf{Dataset} & \textbf{Segment Length} & \textbf{Duration (h)} & \textbf{Input} & \textbf{Model} & \textbf{AUC} & \textbf{ACC} \\
\hline
2011\cite{selvaraj2011statistical} & Multi-site PPG (10 healthy volunteers) & 60s & 1.5 & Kurtosis, Shannon Entropy & Statistical Approach & 0.990 & 0.990 \\\midrule
2012\cite{li2012dynamic} & MIMIC II & 6s & 2.5 & Direct Matching SQI (SQI1), Linear Resampling SQI (SQI2), Dynamic Time Warping SQI (SQI3), Clipping Detection SQI (SQI4) & MLP & 0.952 & 0.975 \\\midrule
2016\cite{elgendi2016optimal} & Custom (heat stress dataset, 40 subjects) & 60s & 1.76 & Perfusion index, skewness, kurtosis, entropy, zero crossing rate, non-stationarity, relative power, matching systolic wave detection & Not mention & F1=0.791 & - \\\midrule
2019\cite{vadrevu2019real} & CSL-PICU, MIT-BIH SLP, MIMIC-II & - & 58.76 & Amplitude, Frequency, Temporal, Derivatives, Composite Features & Rule-Based Decision & 0.978 & 0.953 \\\midrule
2019\cite{pereira2019deep} & Shimmer3 GSR+ & 3s, 6s & 16.03 & 3s-1D PPG/6s-DWT & CNN-LSTM Autoencoder & - & - \\\midrule
\cellcolor[gray]{0.9}\textbf{2019\cite{naeini2019real}} & Empatica E4, PulseOn & 60s & 120 & 1-D PPG signal & 1D CNN & 0.880 & 0.834 \\\midrule
\cellcolor[gray]{0.9}\textbf{2020\cite{goh2020robust}} & Local + Independent (MIMIC II) & 5s & 239.4 & 1-D PPG signal & 1D CNN & - & 0.945 \\\midrule
2020\cite{liu2020classificationA} & Self-made (Impedance Cardiography, ICG) & Pulse & 0.5 & Four features: the number of zero crossings, pulse interval and amplitude differences, and peak smoothness & SoNFIN & - & 0.86 \\\midrule
2020\cite{liu2020classificationB} & Carescape Monitor B650 & Pulsatile & 7.6 & image=PPG + FDP & VGG-19 & - & 0.895 \\\midrule
2021\cite{roh2021recurrence} & VitalDB & 1s & 13.77 & image=Recurrence Plot & 2-D CNN & 0.994 & 0.975 \\\midrule
2021\cite{chen2021signal} & VitalDB & 10s & 16.12 & image=STFT & 2-D CNN & \textbf{0.997} & 0.983 \\\midrule
\cellcolor[gray]{0.9}\textbf{2022\cite{shin2022deep}} & MIMIC III & 5s &\textbf{2214} & 1-D PPG signal & 1D CNN & 0.980 & 0.978 \\\midrule
\cellcolor[gray]{0.9}\textbf{2022\cite{sivanjaneyulu2022cnn}} & MIT-BIH SLP, MIMIC, BIDMC & 3s & 412.07 & 1-D PPG signal & 1D CNN & - & \textbf{0.999} \\\midrule
2022\cite{chatterjee2022signal} & Queensland & 5s & 39.4 & 20x500 Grayscale Images & Lightweight Slim-CNN & 0.992 & 0.983 \\\midrule
\cellcolor[gray]{0.9}\textbf{2023\cite{naeini2023deep}} & UCSF, Neuro ICU & - & 210 & 1-D PPG signal & 1D CNN & 0.962 & 0.952 \\
 &  & - & 210 & 2D image & VGG16 & 0.949 & 0.924 \\
 &  & - & 210 & 2D image & ResNet50 & 0.935 & 0.925 \\
 &  & - & 210 & 2D image & MobileNetV2 & 0.946 & 0.956 \\\midrule
2024\cite{liu2023lightweight} & MIMIC-III, UCI, Queensland & 5s & 99.72 & MMTF transform PPG, FDP, SDP into a 2D image & Hybrid (CNN + Swin Transformer) & 0.930 & 0.934 \\
\midrule
\rowcolor[gray]{0.9} 
\textbf{Proposed} & \textbf{M4M, MIMIC IV} & \textbf{30s} & \textbf{1016.46} & \textbf{3-D signals} & \textbf{LRS-SE} & \textbf{0.964} & \textbf{0.900} \\
\hline
\end{tabular}%
}
\end{table*}

The progression of methodologies in SQA demonstrates a transition from hand-crafted feature extraction to deep learning approaches that obviate the need for manual feature engineering, and more recently, to models that transform signals into image representations for analysis. Similarly, model architectures have advanced from rule-based methods to 1D CNNs and increasingly complex designs. Although lightweight and efficient models are emphasized, most existing studies do not report parameter sizes, making direct comparisons difficult. To address this limitation, we reproduced several SOTA methods highlighted in \textbf{Table III} (indicated by bolded in grey references) in Experiment II to enable a systematic evaluation.

\section{DISCUSSION}

Our experimental results and reproductions of existing studies did not fully replicate the outcomes reported in the original research. This discrepancy is primarily attributed to differences in the datasets. In comparison to the studies summarized in \textbf{Table III}, the majority of referenced research used datasets composed of healthy participants. In contrast, our dataset comprises signals from hospitalized patients experiencing AF episodes. These episodes introduce a complex signal pattern that poses significant challenges to model performance and classification accuracy, substantially increasing the difficulty of SQA. 
Furthermore, our dataset includes over 1,000 hours of long-term 24-hour monitoring data, offering a larger and more diverse set of signals compared to the smaller-scale datasets employed in previous studies, as shown in \textbf{Table III}.
This extensive dataset enables a more comprehensive evaluation of the model's robustness and its applicability in clinically relevant scenarios.

This study compares different input configurations, including PPG, FDP, SDP, and ATC, as well as their various combinations. Previous studies have shown that FDP and SDP are critical for extracting meaningful features. For example, J. Liu et al. \cite{liu2023lightweight} and S.-H. Liu et al. \cite{liu2020classificationB} transformed PPG signals and their derivatives into images for analysis. In contrast, our approach directly processes signals without image transformations, enabling a lightweight network with fewer parameters and lower FLOPs. This design achieves competitive results while reducing model complexity, effectively addressing the high parameter counts and architectural challenges associated with image-based methods.

Results from different comparisons show that if a more lightweight design is desired, using PPG alone as input also exhibits certain advantages, with a 1.24\% improvement on the M4M test set and an 8.34\% improvement on the MIMIC dataset.
This demonstrates that our designed network exhibits adaptability and robustness under different input conditions.

We also conducted experiments using ATC as an input. To analyze its impact, we calculated the average AUC across all input configurations that include ATC. The experimental results in \textbf{Table I} show that for noise classification on the M4M test set, the model with ATC as an input achieved an average AUC of 0.9609 without the SE block and 0.9642 with the SE block, representing an improvement of 0.34\%.
However, performance on the MIMIC dataset decreased: the average AUC was 0.8114 without the SE block, dropping to 0.7171 with the SE block, a decline of 11.62\%.
These results indicate that while the SE block improved the model's performance on the M4M test set, it led to a significant decline in the MIMIC dataset.
A possible explanation is that no training was performed on the MIMIC dataset, and the evaluation was conducted entirely as a cross-dataset validation. This also suggests that the SE block may not generalize well to unseen datasets.

\section{Conclusion}
\label{sec:Conclusion}

This study proposed a lightweight ResNet-based framework with channel attention for efficient PPG SQA. The model demonstrated competitive performance with reduced complexity, making it suitable for wearable devices. Additionally, we conducted a comparison with existing SOTA methods and different preprocessing methods.

\section*{Acknowledgment}

The study was part of a clinical trial, CARE-DETECT (ClinicalTrials.gov ID: NCT05351775), which started on 2022-04-12.

\bibliographystyle{IEEEtran}
\bibliography{mybib}

\end{document}